\documentclass[twocolumn,times]{aastex631}

\usepackage{xspace}
\usepackage[T1]{fontenc}

\newcommand{\kms}{\ensuremath{\mathrm{km\ s^{-1}}}\xspace}
\newcommand{\NH}{\ensuremath{N_{\mathrm{H}}}\xspace}
\newcommand{\cf}{\ensuremath{C_f}\xspace}
\newcommand{\xmm}{{\it XMM-Newton}\xspace}
\newcommand{\iras}{{IRAS~17020+4544}\xspace}
\newcommand{\pg}{{PG~1211+143}\xspace}
\newcommand{\cm}{{\ensuremath{\rm{cm}^{-2}}}\xspace}
\newcommand{\cloudy}{{\tt Cloudy}\xspace}
\newcommand{\lya}{Ly\ensuremath{\alpha}\xspace}
\newcommand{\lyb}{Ly\ensuremath{\beta}\xspace}
\received{2021 November 23}
\accepted{2022 February 16 by ApJ}
\shorttitle{UV counterpart of an X-ray ultra-fast outflow in IRAS 17020+4544}
\shortauthors{Mehdipour et al.}
\graphicspath{{./}{figures/}}
\begin{document}

\title{\Large UV counterpart of an X-ray ultra-fast outflow in IRAS 17020+4544}

\correspondingauthor{Missagh Mehdipour}
\email{mmehdipour@stsci.edu}

\author[0000-0002-4992-4664]{Missagh Mehdipour}
\affiliation{Space Telescope Science Institute, 3700 San Martin Drive, Baltimore, MD 21218, USA}

\author[0000-0002-2180-8266]{Gerard A. Kriss}
\affiliation{Space Telescope Science Institute, 3700 San Martin Drive, Baltimore, MD 21218, USA}

\author[0000-0001-6291-5239]{Yair Krongold}
\affiliation{Instituto de Astronom\'{i}a, Universidad Nacional Aut\'{o}noma de M\'{e}xico, Circuito Exterior, Ciudad Universitaria, Ciudad de M\'{e}xico 04510, M\'{e}xico}

\author[0000-0001-8825-3624]{Anna Lia Longinotti}
\affiliation{Instituto de Astronom\'{i}a, Universidad Nacional Aut\'{o}noma de M\'{e}xico, Circuito Exterior, Ciudad Universitaria, Ciudad de M\'{e}xico 04510, M\'{e}xico}

\author[0000-0001-8470-749X]{Elisa Costantini}
\affiliation{SRON Netherlands Institute for Space Research, Niels Bohrweg 4, 2333 CA Leiden, the Netherlands}
\affiliation{Anton Pannekoek Institute, University of Amsterdam, Postbus 94249, 1090 GE Amsterdam, The Netherlands}

\author[0000-0003-1880-1474]{Anjali Gupta}
\affiliation{Columbus State Community College, 550 E Spring St., Columbus, OH 43215, USA}
\affiliation{Department of Astronomy, The Ohio State University, 140 West 18th Avenue, Columbus, OH 43210, USA}

\author[0000-0002-4822-3559]{Smita Mathur}
\affiliation{Department of Astronomy, The Ohio State University, 140 West 18th Avenue, Columbus, OH 43210, USA}
\affiliation{Center for Cosmology and Astroparticle Physics, 191 West Woodruff Avenue, Columbus, OH 43210, USA}

\author[0000-0002-6896-1364]{Fabrizio Nicastro}
\affiliation{Observatorio Astronomico di Roma-INAF, Via di Frascati 33, 1-00040 Monte Porzio Catone, RM, Italy}

\author[0000-0003-0543-3617]{Francesca Panessa}
\affiliation{INAF - Istituto di Astrofisica e Planetologia Spaziali, via Fosso del Cavaliere 100, I-00133 Roma, Italy}

\author[0000-0002-4814-2492]{Debopam Som}
\affiliation{Space Telescope Science Institute, 3700 San Martin Drive, Baltimore, MD 21218, USA}
\begin{abstract}
We report on the discovery of a UV absorption counterpart of a low-ionization X-ray ultra-fast outflow (UFO) in the Narrow-Line Seyfert-1 galaxy \iras. This UV signature of the UFO is seen as a narrow and blueshifted \lya absorption feature in the far-UV spectrum, taken with the Cosmic Origins Spectrograph (COS) on the {\it Hubble Space Telescope} (HST). The \lya feature is found to be outflowing with a velocity of $-23{,}430$ \kms ($0.078\,c$). We carry out high-resolution UV spectroscopy and photoionization modeling to study the UFO that is seen in the HTS/COS spectrum. The results of our modeling show that the UV UFO corresponds to a low-ionization, low-velocity component of the X-ray UFO found previously with \xmm's Reflection Grating Spectrometer (RGS). The other higher-velocity and higher-ionization components of the X-ray UFOs are not significantly detected in the HST/COS spectrum, consistent with predictions of our photoionization calculations. The multiple ionization and velocity components of the UFOs in \iras suggest a scenario where a powerful primary UFO entrains and shocks the ambient medium, resulting in formation of weaker secondary UFO components, such as the one found in the UV band.
\end{abstract}
\section{Introduction} 
\label{sec:intro}

The observational relations between supermassive black holes (SMBHs) and their host galaxies (e.g. \citealt{Korm13}) suggest that they are co-evolved. However, the mechanism needed for this co-evolution is not yet fully understood. In active galactic nuclei (AGN), accretion onto the SMBH liberates enormous power, but how this power is transferred from the small scales close to the SMBH to the larger scales of the galaxy is uncertain. The accretion in AGN is accompanied by outflowing winds, which transfer energy into the interstellar medium (ISM) of the host galaxy. The resulting feedback mechanism may have an important impact on star formation and galaxy evolution (see e.g. \citealt{Silk98,King15,Gasp17}).

The AGN feedback models suggest the kinetic luminosity of AGN outflows needs to be at least 0.5--5\% Eddington luminosity to have a significant impact on the galaxy evolution \citep{DiMat05,Hopk10}. The energetic {\it ultra-fast outflows} (UFOs, \citealt{Tomb10}), with relativistic outflow velocities, are a crucial component of AGN outflows as they have adequate kinetic luminosity to play a key role in AGN feedback. UFOs consist of highly-ionized outflows, which primarily imprint their absorption signatures in hard X-rays in the Fe-K band, namely through \ion{Fe}{25} and \ion{Fe}{26} lines; see e.g. case studies of \object{PG 1211+143} \citep{Poun03a} and \object{PDS~456} \citep{Reev18a}. X-ray studies of nearby Seyfert-1 galaxies find that about 40\% of them have highly-ionized UFOs, detected in tens of targets \citep{Tomb10,Goff13}. The presence of UFOs alongside molecular outflows have also been discovered, suggesting a physical connection between the small-scale outflows near the accretion disk and the large-scale galactic outflows \citep{Tomb15,Long18,Char20}. There are however still significant gaps in our understanding of the physical structure of UFOs, how they operate in AGN, and what is their relation to the less energetic {warm-absorber outflows} (e.g. \citealt{Blu05,Laha14}), which have moderate outflow velocities ($<$~few 1000~km~s$^{-1}$).

Observational results that have come into light in recent years suggest that absorption signatures of UFOs are not only confined to the hard X-ray band, produced by highly-ionized gas. UFOs with relatively low ionizations and multiple components have been found in the soft X-ray band: Ark~564 \citep{Gupt13}, Mrk~590 \citep{Gupt15}, \iras \citep{Long15}, and Mrk~1044 \citep{Kron21}. Also, counterparts of the hard X-ray UFOs in the Fe-K band have been found in the soft X-ray spectra of PG~1211+143 \citep{Reev18b} and PG~1114+445 \citep{Sera19}. Interestingly, lower-ionization counterparts of the UFOs found in X-rays have been discovered at even lower energies in the UV band. A broad and relativistically blueshifted H~I Ly$\alpha$ absorption feature was found in the HST/COS \citep{Gree12} spectrum of PG~1211+143 by \citet{Kris18a}, matching the velocity of one of the X-ray UFO absorption components. Broad absorption lines (BALs) in PDS~456, namely an unidentified feature short-ward of \lya and a highly blueshifted \ion{C}{4} feature \citep{Hama18}, are also thought to be associated with the X-ray UFO in this AGN. Overall, these recent multi-wavelength findings indicate that UFOs may be multi-ionization outflows in AGN.

High-resolution UV spectroscopy with HST can play a crucial role in UFO studies by observing the UV counterpart of the X-ray UFO, thus providing additional physical information about the outflowing gas, such as their kinematics and multi-ionization structure, which are challenging to ascertain from the X-rays alone. This was demonstrated by the study of the \lya counterpart of the X-ray UFO in PG~1211+143 \citep{Kris18a}. The superior signal-to-noise ratio (S/N) and spectral resolution in the UV band, facilitated thanks to HST/COS, may complement and advance the X-ray UFO studies, which are currently mostly limited to CCD-resolution spectroscopy in the Fe-K band and high-resolution X-ray grating spectroscopy for only the very few brightest AGN. Depending on the photoionization parameters of the UFO gas, its absorption lines may be detectable in the UV band. Such photoionization calculations and UV predictions for UFOs were reported as part of the investigation by \citet{Kris18b}. As long as the outflowing gas is not overly ionized, UV absorption lines such \lya would be produced in the HST/COS energy band. While UFOs are most commonly detected as highly-ionized outflows \citep{Tomb10}, the rarer cases of moderately-ionized X-ray UFOs open a new window for studying these UFOs through high-resolution UV spectroscopy, as previously reported in \pg \citep{Kris18a} and presented here in this paper for \iras.  

\iras is a Narrow-Line Seyfert-1 galaxy at redshift ${z = 0.06040}$ \citep{deGr92} as given in the NASA/IPAC Extragalactic Database (NED). In the \xmm Reflection Grating Spectrometer (RGS, \citealt{denH01}) study of \iras, \citet{Long15} discovered the presence of five UFO components in the soft X-ray band. These UFO components, outflowing with velocities ranging from $-23{,}000$ to $-30{,}000$ \kms, were found along with typical warm-absorber outflows in this AGN. These UFOs in \iras were reported to be capable of playing a significant role in AGN feedback. The modeling of the RGS spectra showed the UFO absorption features are produced by a broad range of low to medium ionization levels, seen mostly as \ion{O}{4} to \ion{O}{8} lines, with also hints of \ion{O}{2} and \ion{O}{3} absorption \citep{Long15}. The modeling of the multi-component UFO features in \iras showed they are present as narrow absorption lines in the RGS spectrum, consistent with the instrumental line broadening. The UFO and the warm absorber in \iras were further studied in a follow-up paper by \citet{Sanf18}. They found significant variations in the velocity and ionization of two components of the warm absorber, seen between the 2004 and 2014 \xmm observations. The variable warm-absorber was interpreted as a consequence of a `shocked outflow' scenario, where the UFO sweeps and shocks the ambient medium. Most recently, \citet{Salo21} studied the molecular outflows in \iras in the context of AGN feedback, which may possibly have an association to the observed UFO in this AGN. Interestingly, \iras is one of a few AGN showing compelling evidence for an energy-conserving outflow on galactic scales. This was first discovered by \citet{Long18} using observations with the Large Millimeter Telescope (LMT), and since then has been verified (Longinotti et al. in prep) using observations with the NOrthern Extended Millimeter Array (NOEMA). The radio observations of \iras suggest the presence of a jetted non-thermal source at milli-arcsecond spatial scales \citep{Giro17}; the inter-play between this radio source and the outflowing wind is currently under investigation (Stanghellini et al. in prep).

In this paper we investigate the UV spectral signatures of the X-ray UFOs in \iras using our HST/COS observations taken in 2018. This enables us to examine the multi-ionization structure of the UFOs in \iras by exploring the presence of any expected UV counterparts of the X-ray UFOs, and accurately ascertain the parameters of the outflows thanks to the high-resolution and high-S/N HST/COS data. Since UV counterparts of X-ray UFOs have only been found and studied in a couple of AGN (PG~1211+143, \citealt{Kris18a}; PDS 456, \citealt{Hama18}), currently a general understanding of the relation between UV and X-ray components of the UFOs is missing. By investigating this in a new AGN, and thereby increasing the number of cases, the differences and similarities between the multi-component UFOs in AGN can be investigated. This helps towards delineating a general physical picture of the UFO phenomenon in AGN.

The structure of the paper is as follows. In Section \ref{sect_data} we describe the HST/COS observations of \iras and their data processing and preparation. The spectroscopic analysis and modeling of the UFO as seen in the HST/COS spectrum are presented in Section \ref{sect_model}. In Section \ref{sect_discuss} we discuss and interpret the results of our modeling and give our conclusions. Our HST/COS spectral fitting and photoionization modeling in this paper were done in {\tt Python} and with the {\tt Cloudy} v17.02 photoionization code \citep{Ferl17}.

\section{Observations and data processing} 
\label{sect_data}

The log of the HST/COS observations of \iras, proposed by our team (PI: Krongold) and observed in HST Cycle 25, is provided in Table \ref{table_log}. The observations, taken over 9 orbits during COS Lifetime Position 4 (LP4), used grating G140L with the central wavelength setting of 1105~\AA, covering the wavelength range of 1101--2291 \AA. The COS exposures were taken at all four grating offset positions (FP-POS), so that the spectrum falls on slightly different areas of the detector, thus eliminating the effects of detector artifacts such as flat-field features \citep{Hirs21}.

Our HST/COS data of \iras were retrieved from the Mikulski Archive for Space Telescopes (MAST) and processed with the latest COS calibration pipeline, CalCOS v3.3.10. These data can be accessed via \dataset[10.17909/gpc6-xv04]{https://doi.org/10.17909/gpc6-xv04}. We verified the expected accuracy of the wavelength calibration by examining the observed wavelengths of suitable Galactic ISM lines in the spectrum. All COS exposures were combined to produce one calibrated merged spectrum. This COS spectrum was then binned by four pixels to improve the S/N while still oversampling the 10-pixel resolution element on the far-UV detector \citep{Fox18}. An overview of the final COS spectrum of \iras in the far-UV band is shown in Figure \ref{fig_overview}.
\smallskip

%
\begin{deluxetable}{c c c c}
\tablecaption{Log of our HST/COS observations of \iras taken over 9 orbits.	
\label{table_log}}
\tablewidth{0pt}
\tablehead{
\colhead{Instrument\,/} & \colhead{Dataset} & \colhead{Start time} & \colhead{Exposure}\\
\colhead{Grating} & \colhead{ID} & \colhead{yyyy-mm-dd hh:mm:ss} & \colhead{(ks)}
}
\startdata
COS\,/\,G140L & LDHP01010	& 2018-11-23 22:50:35 & 7.61 \\
COS\,/\,G140L & LDHP02010	& 2018-11-24 03:43:35 & 7.61 \\
COS\,/\,G140L & LDHP03010	& 2018-11-24 08:29:31 & 7.61 \\
\enddata
\tablecomments{The data correspond to proposal ID number 15239, obtained in HST Cycle 25. All G140L exposures were taken with the central-wavelength setting 1105~\AA.}
\end{deluxetable}

\section{Spectral analysis and modeling} 
\label{sect_model}

We began our spectral analysis of \iras by examining regions of the COS spectrum where any UV counterpart of the X-ray UFO components may appear. The best-fit parameters of the X-ray UFO, measured most recently by \citet{Sanf18}, are provided in Table \ref{table_xray}. The X-ray UFO is found to have four ionization components, named `UFO 1', `UFO 2', `UFO 3', and `UFO 4'. However, the X-ray UFO consists of only two velocity components, which we refer to as the `high-velocity' (UFO 1 and UFO 2) and `low-velocity' (UFO 3 and UFO 4) components of the X-ray UFO. Only the UFO 1 and UFO 3 components are persistently present in both the 2004 and 2014 X-ray data. The UFO 2 and UFO 4 components are not seen in the 2004 data, and thus may not always be present in our line of sight.

\begin{figure*}
\centering
\resizebox{\hsize}{!}{
\includegraphics[angle=0]{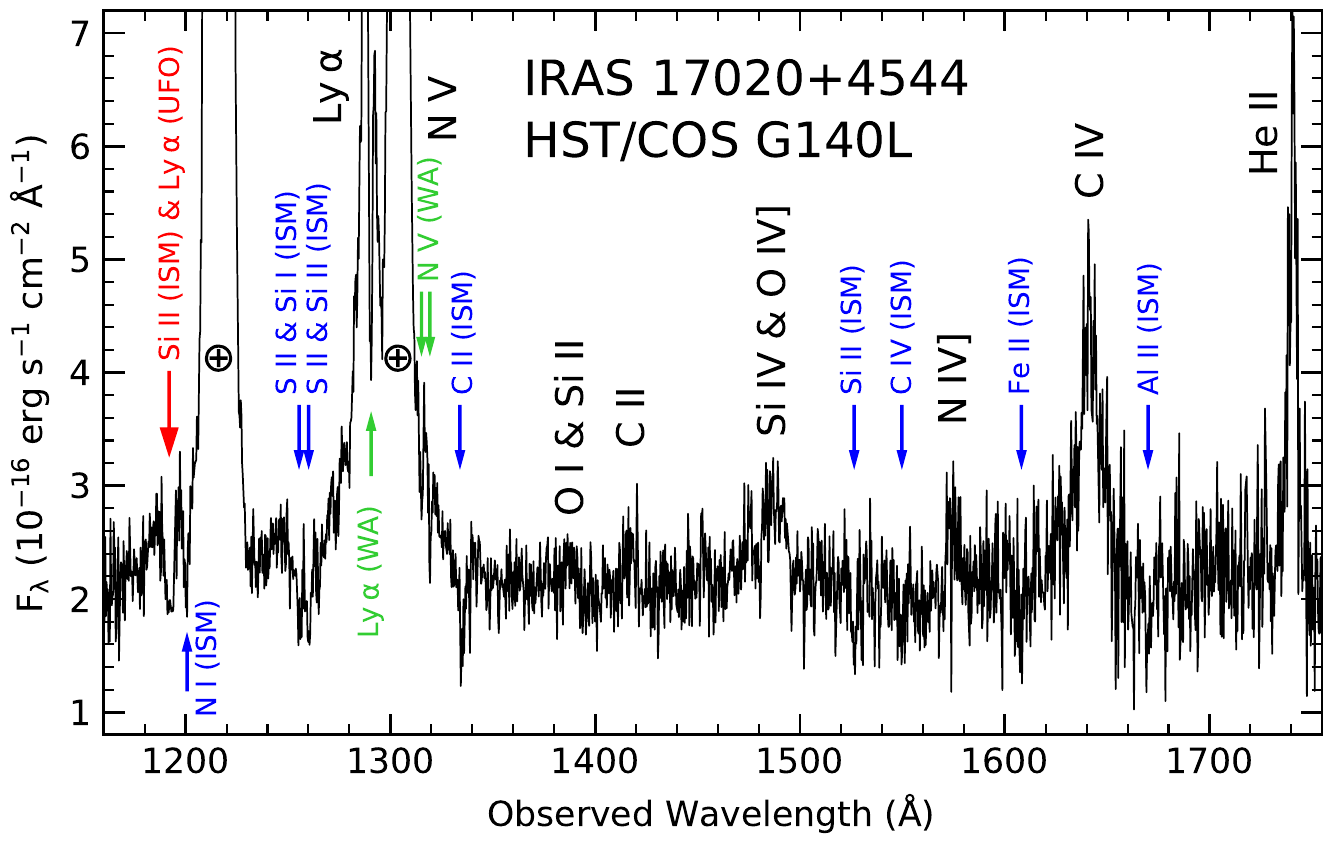}
}
\caption{An overview of the HST/COS far-UV spectrum of \iras taken with the G140L grating. The AGN emission lines are labeled in black. The ISM absorption lines are shown with blue arrows and labels, which in the order of increasing wavelength are: \ion{Si}{2} doublet ($\lambda$1190, $\lambda$1193), \ion{N}{1} ($\lambda$1200), \ion{S}{2} triplet ($\lambda$1251, $\lambda$1254, $\lambda$1260), \ion{Si}{1} ($\lambda$1255), \ion{Si}{2} ($\lambda$1260), \ion{C}{2} ($\lambda$1335), \ion{Si}{2} ($\lambda$1527), \ion{C}{4} doublet ($\lambda$1548, $\lambda$1551), \ion{Fe}{2} ($\lambda$1608), and \ion{Al}{2} ($\lambda$1671). The intrinsic absorption lines from the warm absorber (WA) of the AGN are shown with green arrows and labels, which are \lya ($\lambda$1216) observed at 1290.6 \AA, and the \ion{N}{5} doublet ($\lambda$1239, $\lambda$1243) observed at 1315.2 \AA\ and 1319.4 \AA. The geocoronal emission lines are indicated with the Earth symbol $\oplus$. The red arrow shows the location of the \lya UFO absorption feature at 1192.0 \AA\ and the nearby Galactic \ion{Si}{2} ISM lines in the COS spectrum. A close-up of this region, and our best-fit model to the \lya and the ISM lines, are shown in Figure \ref{fig_ufo}. The blueshifted \lya feature is found to be the UV spectral counterpart of the low-ionization, low-velocity component of the X-ray UFO in this AGN.
\label{fig_overview}}
\vspace{0.3cm}
\end{figure*}

\begin{figure*}
\centering
\resizebox{\hsize}{!}{
\includegraphics[angle=0]{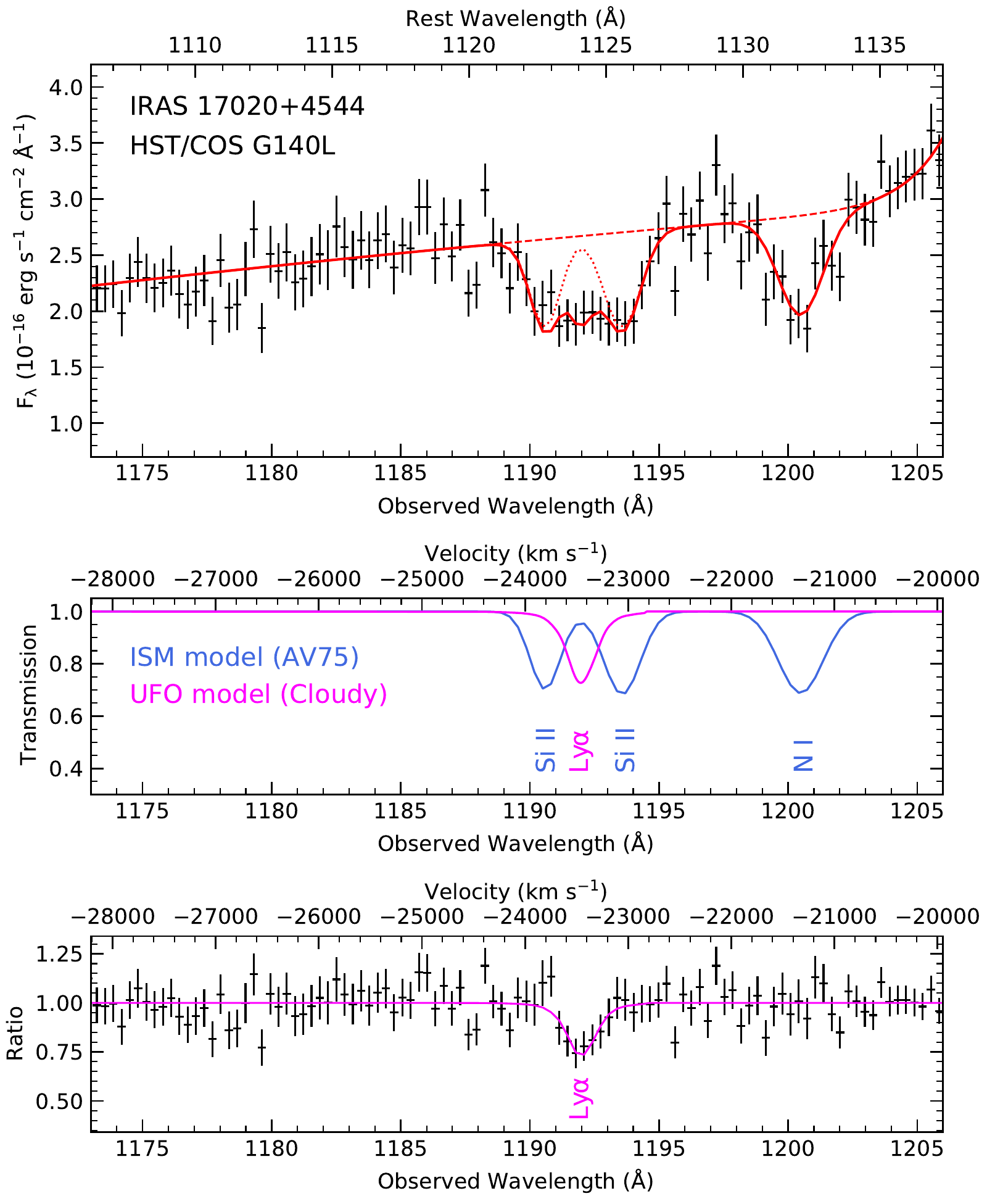}
\includegraphics[angle=0]{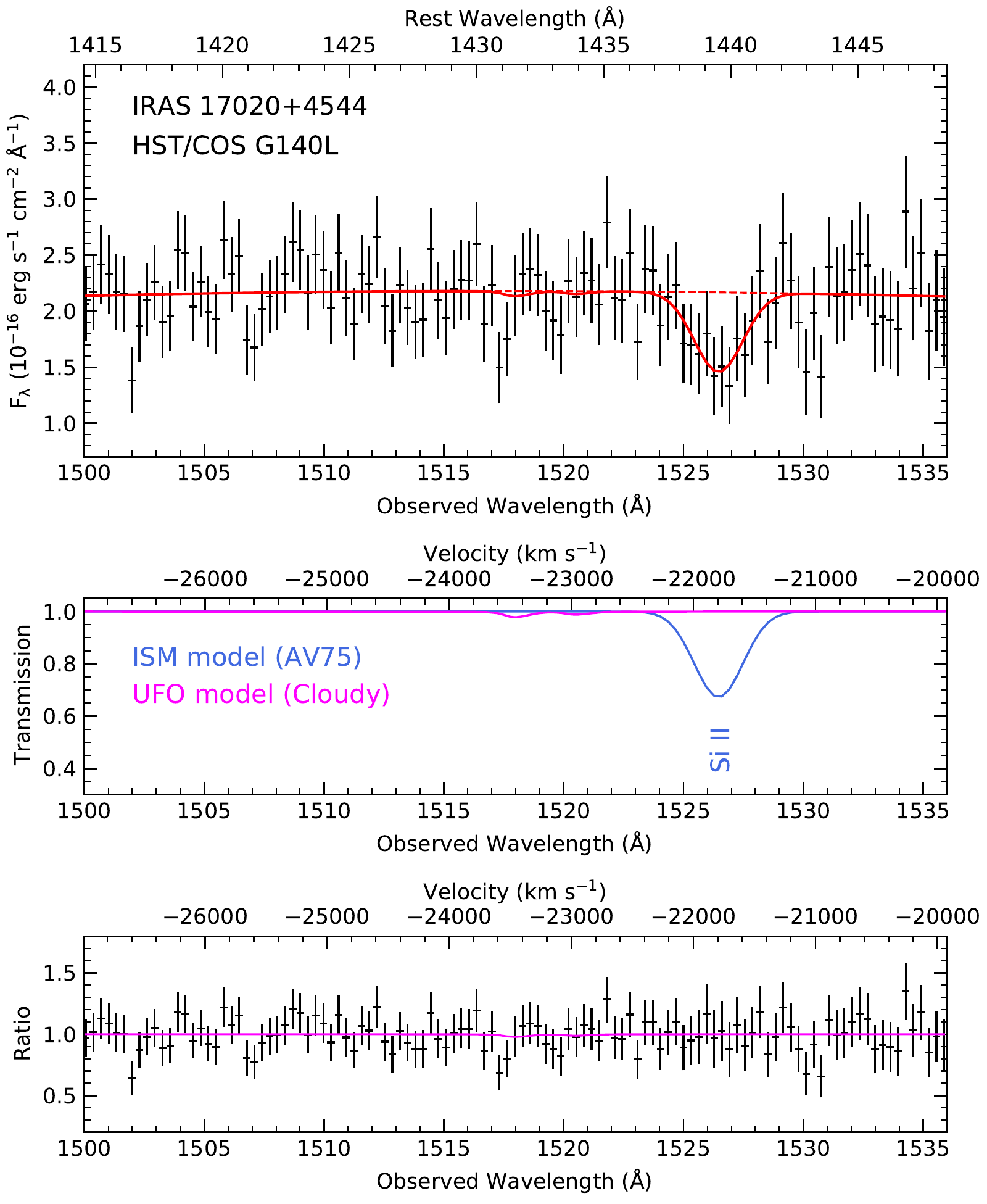}
}
\caption{{\it Left panels}: \lya absorption counterpart of a low-ionization X-ray UFO component in \iras. This narrow and blueshifted \lya feature, found in the HST/COS spectrum, is outflowing with a velocity of $-23{,}430$~\kms. In the {\it top left panel}, the best-fit model to the \lya feature, and the adjacent Galactic ISM lines, is shown in solid red line; the dotted red line denotes the best-fit model without including the \lya absorption line; and the continuum model is shown as a dashed red line. {\it Right panels}: The predicted absorption model in the \ion{C}{4} region, corresponding to the model fitted to the \lya feature. The calculated UFO model does not produce any significant \ion{C}{4} absorption, consistent with the observed HTS/COS spectrum. {\it Middle panels}: Transmission of the best-fit model for the UFO (computed with \cloudy) and the ISM (derived from the COS calibration star AV75 in Figure \ref{fig_ism}), shown in the \lya and \ion{C}{4} regions. {\it Bottom panels}: Ratio of the data to the model that excludes the UFO absorption, showing the presence of the \lya feature in the fit residuals. The S/N of the \lya feature is 6\,$\sigma$. The magenta line represents the best-fit model that includes the UFO absorption divided by the model that excludes the UFO absorption.
\label{fig_ufo}}
\vspace{0.4cm}
\end{figure*}

By taking into account the outflow velocities of the X-ray UFO components (Table \ref{table_xray}), and the cosmological redshift of the AGN, we calculated the resulting observed wavelengths of typical UV absorption lines, such as \lya and \ion{C}{4}, and searched for them in the COS spectrum. Remarkably, we found a \lya absorption feature corresponding to the low-velocity components of the X-ray UFO (Table \ref{table_xray}), which we will give more detail on later in Section \ref{sect_ufo}. The location of this \lya feature, outflowing with a velocity of $-23{,}430$~\kms, is indicated by the red arrow in the overview spectrum of Figure \ref{fig_overview}. A close-up of this UFO region is presented later in Figure \ref{fig_ufo}. Incidentally, the blueshifted \lya UFO feature, observed at a central wavelength of 1192.0 \AA, happens to fall in a region of the spectrum where there are also two foreground Galactic ISM lines: the \ion{Si}{2} doublet at $\lambda$1190.4 and $\lambda$1193.3.

%
\begin{deluxetable}{c c c c}
\tablecaption{Parameters of the X-ray UFO components in \iras, obtained by \citet{Sanf18} from 2014 \xmm observations.
\label{table_xray}}
\tablewidth{0pt}
\tablehead{
\colhead{Component}   & \colhead{$\log U$}     & \colhead{\NH}                & \colhead{$v_{\rm out}$}\\
\colhead{name}        & \colhead{}             & \colhead{($10^{20}$ \cm)}    & \colhead{(\kms)}
}
\startdata
UFO 1 & $-2.5 \pm 0.2$  & $1.3 \pm 0.2$    & $-26{,}900 \pm 200$ \\
UFO 2 & $+2.6 \pm 0.1$  & $5000 \pm 2000$  & $''$ \\
UFO 3 & $-0.4 \pm 0.2$ & $2.5 \pm 1.5$    & $-24{,}100 \pm 100$ \\
UFO 4 & $-1.1 \pm 0.1$  & $7 \pm 1$        & $''$ \\
\enddata
\tablecomments{The modeling adopts full covering fraction (${\cf = 1}$). Only the UFO 1 and UFO 3 components have been persistently present, as UFO 2 and UFO 4 were not seen in earlier 2004 X-ray data. We refer to UFO 1 and UFO 2 as the `high-velocity' components, and UFO 3 and UFO 4 as the `low-velocity' components of the X-ray UFO. The low-velocity components of the X-ray UFO can alternatively be modeled as one component with ionization parameter ${\log U = -0.4 \pm 0.3}$, column density ${\NH = 3 \pm 1 \times 10^{21}}$ \cm, and the outflow velocity ${v_{\rm out} = -23{,}640 \pm 100}$ \kms (`Comp. A' in \citealt{Long15}).
}
\end{deluxetable}

Before considering the feature at 1192.0~\AA\ as the UV counterpart of the X-ray UFO, we first investigated whether or not it can be a line from the warm absorber of \iras, or any intervening gas between us and \iras. Using the X-ray warm absorber model of \citet{Sanf18}, we calculated the corresponding UV absorption model and compared that with our HST/COS spectrum. The X-ray warm-absorber consists of four components, with their velocities ranging from an inflow of ${+1{,}750 \pm 250}$ \kms to an outflow of ${-2{,}300 \pm 200}$ \kms. Our UV calculations were done using the \cloudy code with the parameters of the warm-absorber components taken from Table 1 of \citet{Sanf18}. From this comparison between the model and the COS spectrum we ascertain that the feature at 1192.0~\AA\ cannot be attributed to any neutral or ionized absorption line from the warm absorber (see also Sect. \ref{sect_ref}). Our analysis of the COS spectrum shows that the only UV warm-absorber present, seen as \lya and \ion{N}{5} lines (Figure \ref{fig_overview}), has an inflow velocity of about ${+350 \pm 50}$~\kms, which corresponds to the `WA 1' component of \citet{Sanf18} at ${+320 \pm 70}$~\kms, as well as the molecular CO line found by \citet{Long18} at $+233$~\kms (their `Line B' component). This intrinsic UV absorption component, which has a relatively low velocity, cannot explain the highly blueshifted feature at 1192.0 \AA. Our calculations show no neutral or ionized absorption line at this wavelength can be produced by the UV warm absorber considering the low velocity and the column density of the observed \lya and \ion{N}{5} lines. Therefore, we deduce that the feature at 1192.0 \AA\ is not a line from any X-ray or UV component of the warm absorber in \iras. The lack of any significant UV absorption by the other components of the X-ray warm-absorber in the COS spectrum is likely because of the ionization state of those components and/or their low covering fraction of the UV source. Our team plans to publish a separate paper on the UV warm-absorber in \iras and its relation to the molecular and X-ray outflows in this AGN. Hence, while we have established that the 1192.0 \AA\ feature cannot be any warm-absorber line, we do not discuss the warm absorber further in this paper, which is focused on the study of the UFO in \iras.

%

The feature at 1192.0~\AA\ cannot be from any intervening intergalactic medium (IGM). Any feasible IGM line would only appear at wavelengths ${> 1216}$~\AA\ produced by \lya, whereas the feature is observed at 1192.0~\AA. Likewise, there is no possible interstellar medium (ISM) line at the wavelength of 1192.0~\AA; the modeling of the actual ISM lines near the feature is shown in Section \ref{sect_ism}. The 1192.0~\AA\ feature cannot be explained by any ion or line other than \lya, because otherwise various associated lines would also be produced, which are not observed in the COS spectrum. This point is demonstrated by the series of spectral calculations provided later in Section \ref{sect_ref} (Figure \ref{fig_ref}). The feature at 1192.0~\AA\ can only be explained as a blue-shifted \lya line, produced by a gas with high outflow velocity. Since such a high outflow velocity gas is already seen in \iras as a UFO in the X-ray band \citep{Long15,Sanf18}, and the velocity of the \lya feature is consistent with that of the low-ionization, low-velocity component of the X-ray UFO, it is reasonable to infer that the feature is associated to the X-ray UFO. We investigate this by photoionization modeling and spectral fitting in Section \ref{sect_ufo}. 

The blueshifted \lya line at 1192.0~\AA\ is blended with the two adjacent \ion{Si}{2} lines, therefore, it is essential to model precisely both the UFO and ISM lines. Importantly, COS G140L is capable of resolving the two lines of the \ion{Si}{2} doublet, as we demonstrate in Section \ref{sect_ism}, where we derive an ISM model from G140L observations of other targets with the same instrumental settings. We then apply our derived ISM model to the COS spectrum of \iras in Section \ref{sect_ufo}, where we carry out parametrization and photoionization modeling of the UV UFO feature.

We note that in this paper the outflow velocity that we quote and show in figures is the relativistic outflow velocity $v_{\rm out}$, calculated as \newline
$v_{\rm out} = c\, [(1 + z_{\rm out})^2 - 1]\, /\, [(1 + z_{\rm out})^2 + 1]$, \\
where $c$ is the speed of light, \newline 
$z_{\rm out} = [(\lambda_{\rm obs}/\lambda_{0})\, /\, (1 + z)] - 1$, \newline
$\lambda_{\rm obs}$ is the observed wavelength of a spectral line, and $\lambda_{0}$ is the rest wavelength of the spectral line, and $z$ is the cosmological redshift. For \iras ${z = 0.06040}$ according to optical spectroscopy by \citet{deGr92}, or alternatively ${z = 0.0612}$ based on the recent study of the double-peak CO lines by \citet{Salo21}. In this paper we adopted the redshift value which was used in the prior X-ray studies of \iras \citep{Long15,Sanf18}, which is ${z = 0.06040}$, for a consistent comparison with the velocities of the X-ray absorption components. Nonetheless, the velocity difference between the two redshift values ($225$~\kms) is relatively small compared to the velocity of the UFO in \iras.

%

\subsection{Modeling of the ISM absorption lines}
\label{sect_ism}

To model the blueshifted \lya feature of the UFO in \iras, a model for the neighboring ISM lines in the HST/COS spectrum is required. We derived this model using HTS/COS G140L observations of an O-type star named \object{AV75}, which has been a standard reference target for calibrating the resolution of the COS instrument \citep{Fox18}. For comparison and further checks, we also produced a mean ISM spectrum from HST/COS G140L observations of those quasi-stellar objects (QSOs) with the highest S/N for spectroscopy of the \ion{Si}{2} and \ion{N}{1} ISM lines, that were taken with identical COS instrumental setup and settings as our \iras observations. This resulted in selection of seven QSOs. The selected AV75 and QSO datasets\footnote{Dataset IDs: LDQ701070, LDUH01070, LDUH51070, LE0R01070,\newline LEFC01070, LE2J01010, LE2J02010, LE2J03010, LE2J05010, \newline LE2J06010, LE2J12010, LE2J13010.} were observed with the G140L grating, the central-wavelength setting of 1105~\AA, taken during the COS LP4 period, which are the same as our \iras observations. Therefore, the ISM transmission model that is derived, which is already convolved with the instrumental effects such as the line spread function (LSF), is directly applicable to the COS spectrum of \iras. The processing of the AV75 and QSO data follows the same procedure described in Section \ref{sect_data} for \iras.

\begin{figure}
\centering
\resizebox{\hsize}{!}{
\includegraphics[angle=0]{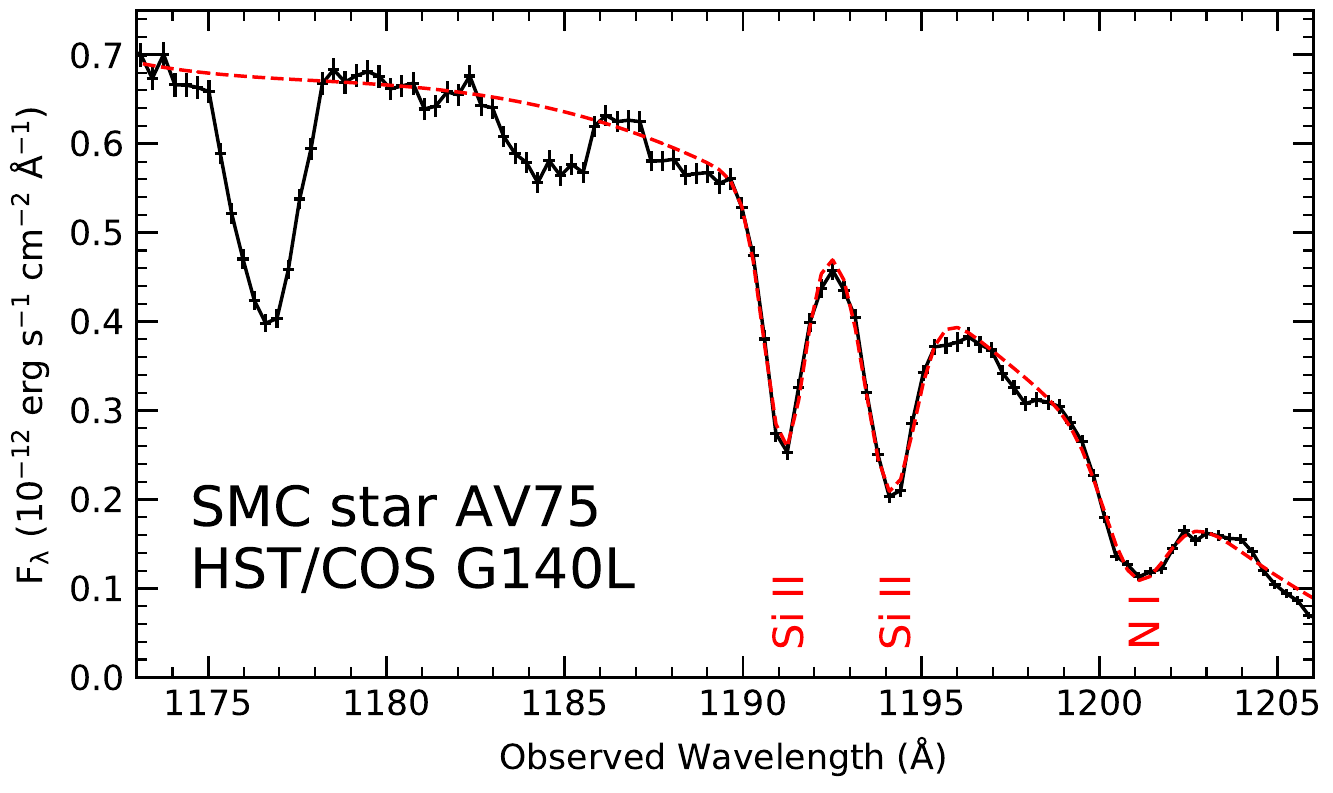}
}
\resizebox{\hsize}{!}{
\includegraphics[angle=0]{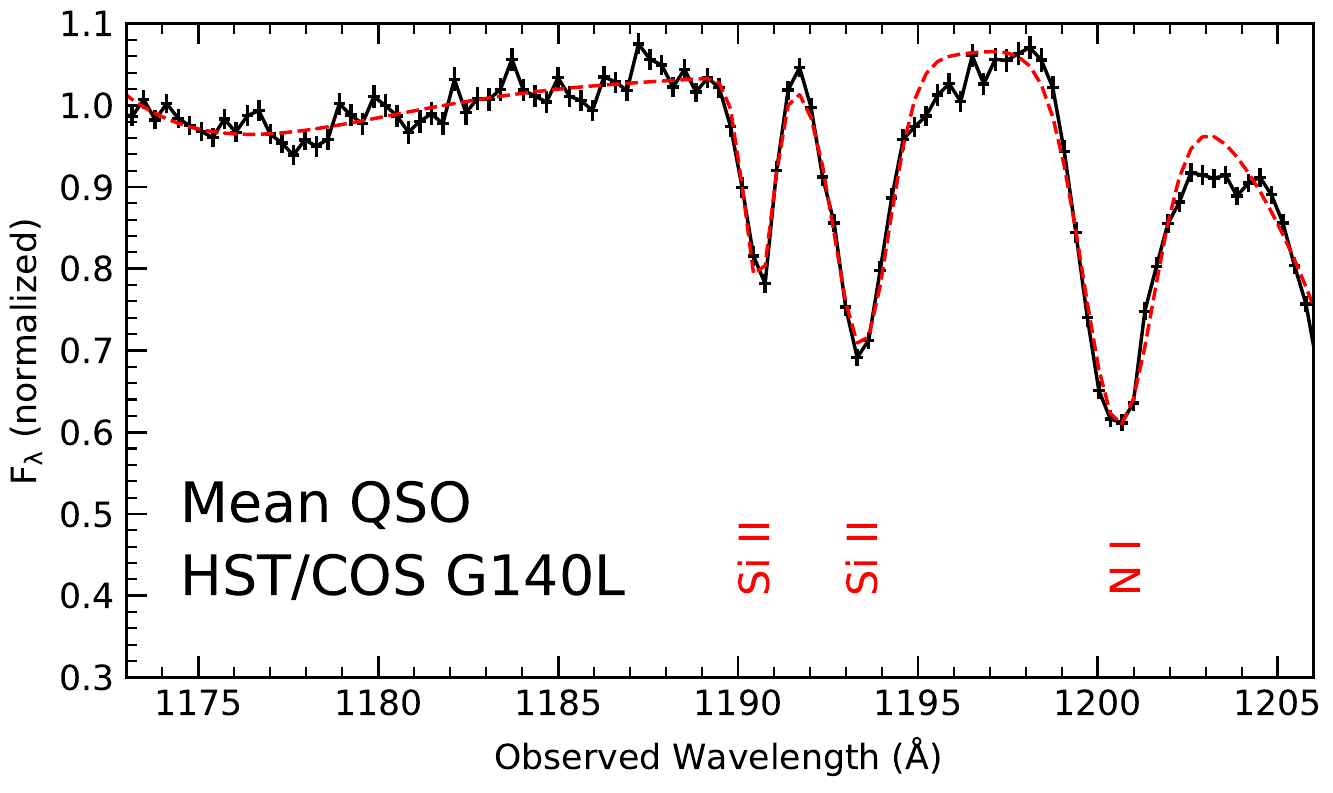}
}
\caption{HST/COS G140L spectra of the calibration star AV75 ({\it top panel}) and the stacked QSO sample ({\it bottom panel}), displayed in the region where the \lya UFO feature is found in \iras (Figure \ref{fig_ufo}, {\it left panels}). The best-fit model to the ISM \ion{Si}{2} and \ion{N}{1} lines is shown as a dashed red line. The \ion{Si}{2} doublet lines are clearly resolved by G140L in the AV75 and QSO spectra, and unlike \iras there is no additional absorption feature present between the two \ion{Si}{2} lines.
\label{fig_ism}}
\end{figure}

Figure \ref{fig_ism} shows the HTS/COS G140L spectrum of AV75 (top panel) and the stacked QSO sample (bottom panel). Our best-fit model to the ISM lines of the \ion{Si}{2} doublet and \ion{N}{1} are also displayed. For modeling these narrow ISM lines a local continuum fit is adequate as here we are not concerned with the intrinsic broadband continuum of the objects. To this end, we fitted a polynomial function to represent both the continuum and any other underlying broad spectral features. This local continuum model is fitted over the band displayed in Figure \ref{fig_ism}, while excluding wavelengths where any narrow feature resides. Each of the three absorption lines (\ion{Si}{2} doublet and \ion{N}{1}) was fitted with a Gaussian function as shown in Figure \ref{fig_ism}. As AV75 is in the Small Magellanic Cloud (SMC), its observed stellar absorption lines are redshifted by $+145$~\kms. This velocity shift is taken into account in our modeling of the lines.
 
%
 
The ISM transmission model derived from AV75 is used for our final fitting of the \iras spectrum in Section \ref{sect_ufo} (Figure \ref{fig_ufo}). The AV75 ISM model is normalized to take into account the Galactic column density in our line of sight towards \iras (${\NH = 2.0 \times 10^{20}}$~\cm, \citealt{Kal05}) while maintaining the line ratios and widths according to the AV75 ISM model. Since AV75 is the reference calibration target for HST/COS, it is the most appropriate target to use. In any case, we find that the results of our UFO modeling remains the same by using the ISM model derived from the mean QSO spectrum. The individual ISM lines in the AV75 and QSO spectra (Figure \ref{fig_ism}) are narrow as their widths are consistent with predominantly instrumental broadening by G140L. Importantly, both the AV75 and QSO spectra show that the individual lines of the \ion{Si}{2} doublet are resolved (Figure \ref{fig_ism}), and there is no additional feature present between the two \ion{Si}{2} lines. This is in contrast to the \iras spectrum, where there is an additional absorption feature (Figure \ref{fig_ufo}, left panels), corresponding to the \lya absorption line of the X-ray UFO. The fact that the \ion{N}{1} line in the \lya region (Figure \ref{fig_ufo}, left panels) and the \ion{Si}{2} line in the \ion{C}{4} region (Figure \ref{fig_ufo}, right panels) are both narrow and fitted well with the AV75 ISM model, verifies that the \ion{Si}{2} doublet lines that are around the \lya feature must also be narrow. Therefore, two peculiarly broad \ion{Si}{2} lines cannot explain the blended feature at 1192.0 \AA\ as they would be physically inconsistent with the other ISM lines in the COS spectrum. Our modeling of the UV UFO in \iras is described below in Section \ref{sect_ufo}.

\subsection{Modeling of the UV UFO in \iras}
\label{sect_ufo}

We used the \cloudy v17.02 code \citep{Ferl17} for both photoionization modeling and the calculation of the UV spectrum of the UFO in \iras. We produced a grid of model spectra computed by \cloudy to fit the \lya feature (Figure \ref{fig_ufo}, left panels), and to also calculate and examine the corresponding UV absorption model in other regions of the COS spectrum, notably the \ion{C}{4} doublet (Figure \ref{fig_ufo}, right panels) where the second strongest signature would be detectable. The associated blueshifted \ion{N}{5} doublet falls on the geocoronal \lya emission line at 1216 \AA\ and thus is unobservable. Also, any corresponding blueshifted \lyb and \ion{O}{6} lines from the UFO would fall outside of the COS spectral band.

For photoionization modeling we used one of the built-in spectral energy distributions (SEDs) in \cloudy that represents the SED of a typical type-1 AGN. This SED corresponds to the intrinsic broadband continuum of NGC~5548, derived by \citet{Meh15a} from an extensive multi-wavelength campaign. The choice of the specific SED does not affect our finding of the \lya UFO feature in the COS spectrum, but rather for a given ionization parameter $U$ the inferred column density \NH that fits the data can be slightly different if adopting other type-1 AGN SED models. The relation between the ionization parameter $U$ (defined by \citealt{Davi77}) and the ionization parameter $\xi$ (defined by \citealt{Kro81}) is: ${\log U = \log \xi - 1.62}$. For our \cloudy calculations we adopted the proto-solar abundances of \citet{Lod09}.

As the X-ray lines of the UFO in \iras were seen to be narrow (consistent with instrumental broadening) in the RGS spectrum \citep{Long15,Sanf18}, in our \cloudy calculations we did not add any turbulent line broadening. The calculated model spectra from \cloudy, which contain natural and thermal line broadenings, were then convolved with the appropriate COS Line Spread Function (LSF) for our observations to take into account the instrumental line broadening. We find the resulting \lya profile matches well the observed UFO feature in the COS spectrum without needing any additional line broadening. The \lya UFO feature (Figure \ref{fig_ufo}, left panels) is consistent with being narrow (i.e. consistent with instrumental broadening) in the HST/COS spectrum.

For fitting the observed local continuum we used a polynomial function to take into account both the intrinsic continuum and any contribution from broad spectral features, such as the wing of the nearby geocoronal \lya emission line. In our modeling we also incorporated the ISM transmission model that we derived earlier in Section \ref{sect_ism} to fit the narrow ISM lines of \ion{Si}{2} and \ion{N}{1}.

In our modeling of the \lya feature we fixed the ionization parameter $U$ to that of the persistent low-ionization, low-velocity component of the X-ray UFO (Table \ref{table_xray}): ${\log U \sim -0.4}$, i.e. `UFO 3' in \citet{Sanf18} or `Comp. A' in \citet{Long15}. As the outflow velocity of the blueshifted \lya feature ($-23{,}430$ \kms) is consistent with that of this low-velocity X-ray UFO component (${\sim -23{,}640}$ \kms in \citealt{Long15}, or ${\sim -24{,}100}$ \kms in \citealt{Sanf18}), it is reasonable to assume they originate from similar photoionized regions. Indeed our adopted ionization parameter ${\log U = -0.4}$ is fully consistent with the observed COS spectrum, where only \lya is significantly detected, and other lines such as the \ion{C}{4} and \ion{N}{5} doublets are not present in the spectrum (see Figures \ref{fig_overview}, \ref{fig_ufo}, and \ref{fig_ref}). For the low-velocity X-ray UFO components (Table \ref{table_xray}), the \cloudy predictions show that, compared to the 1192 \AA\ \lya feature, absorption lines from other ions would be too weak to be significantly detected in the COS spectrum (Figures \ref{fig_ufo} and \ref{fig_ref}).

%
\begin{deluxetable}{c | c c c c}
\tablecaption{Best-fit parameters of our modeling of the Ly$\alpha$ UFO feature found at 1192 \AA\ in the HST/COS spectrum of \iras (Figure \ref{fig_ufo}). 
\label{table_para}}
\tablewidth{0pt}
\tablehead{
 & \colhead{$\NH$} & \colhead{\cf} & \colhead{$\log U$} & \colhead{$v_{\rm out}$} \\
Model & \colhead{($10^{20}\cm$)} &  &  & \colhead{($\kms$)}
}
\startdata
A	&	$1.0 \pm 0.2$ & $1$ (f) & $-0.4$ (f) & $-23{,}430 \pm 50$ \\
B	&	$30$ (f) & $0.65 \pm 0.10$ & $-0.4$ (f) & $-23{,}430 \pm 50$ \\
\enddata
\tablecomments{Our Model A, where the UFO gas fully covers the UV source, is consistent with the persistent low-velocity component of the X-ray UFO (UFO 3 in Table \ref{table_xray}). Our Model B, where the UFO partially covers the UV source with higher column density, corresponds to both of the low-velocity components of the X-ray UFO (UFO 3 and UFO 4 in Table \ref{table_xray}). The `(f)' denotes the parameter is fixed. The parameter errors are given at the 1\,$\sigma$ confidence level. The best-fit statistics for both models are ${\chi^2 = 93}$ for 106 degrees-of-freedom (DOF).}
\end{deluxetable}

We fitted the blueshifted \lya absorption feature at 1192 \AA\ (Figure \ref{fig_ufo}, left panels) by parameterizing the column density (\NH), covering fraction (\cf), and the outflow velocity ($v_{\rm out}$). We considered two possible scenarios: Model A, where full covering of the UV source is assumed (${\cf = 1}$) and \NH is fitted; Model B, where \cf is fitted, while \NH is fixed to the total \NH of the low-velocity components of the X-ray UFO (UFO 3 and UFO 4 in Table \ref{table_xray}), which \citet{Long15} have modeled them as one component with ${\NH = 3.0 \times 10^{21}}$~\cm and ${\log U = -0.4}$. Since UFO 3 and UFO 4 have identical velocities, their \lya absorption lines would be blended and thus cannot be distinguished in our modeling, therefore we model UFO 3 and UFO 4 as one component like in \citet{Long15}.

Both Model A and Model B provide equally good fits to the COS data and look almost identical to each other. Interestingly, our fitted parameters of Model A nearly match within errors the parameters of the persistent low-velocity component of the X-ray UFO (UFO 3 in Table \ref{table_xray}). Our best-fit parameters of Model A and Model B are given in Table \ref{table_para}. The best-fit model to the \lya feature is shown on the left panels of Figure \ref{fig_ufo}, and its predicted absorption model in the \ion{C}{4} region (which is insignificant) is shown on the right panels of Figure \ref{fig_ufo}. We use Model A for display in Figure \ref{fig_ufo} (Model B would look almost identical). We discuss the modeling results in Section \ref{sect_discuss}.

The blueshifted \lya absorption line at 1192.0 \AA\ has an equivalent width (EW) of 0.42 \AA\ and is a significant feature in the COS spectrum (S/N = 6\,$\sigma$). By including the \lya line in our modeling, the best fit is improved from ${\chi^2 = 128}$ (${{\rm DOF} = 108}$) to ${\chi^2 = 93}$ (${{\rm DOF} = 106}$). This ${\Delta \chi^2 = 35}$ is statistically highly significant. The corresponding F-test statistic value is 19.9 with a p-value of $4 \times 10^{-8}$. The significance of the \lya feature is also evident by the fact that the fitted model parameters are constrained (relatively small errors in Table \ref{table_para}). On the other hand, the inclusion of blue-shifted \ion{C}{4} absorption (Figure \ref{fig_ufo}, right panels) makes no significant difference to the fit in that region of the spectrum: ${\chi^2 = 69}$ (${{\rm DOF} = 110}$) to ${\chi^2 = 66}$ (${{\rm DOF} = 108}$). 

\begin{figure}
\centering
\resizebox{\hsize}{!}{
\includegraphics[angle=0]{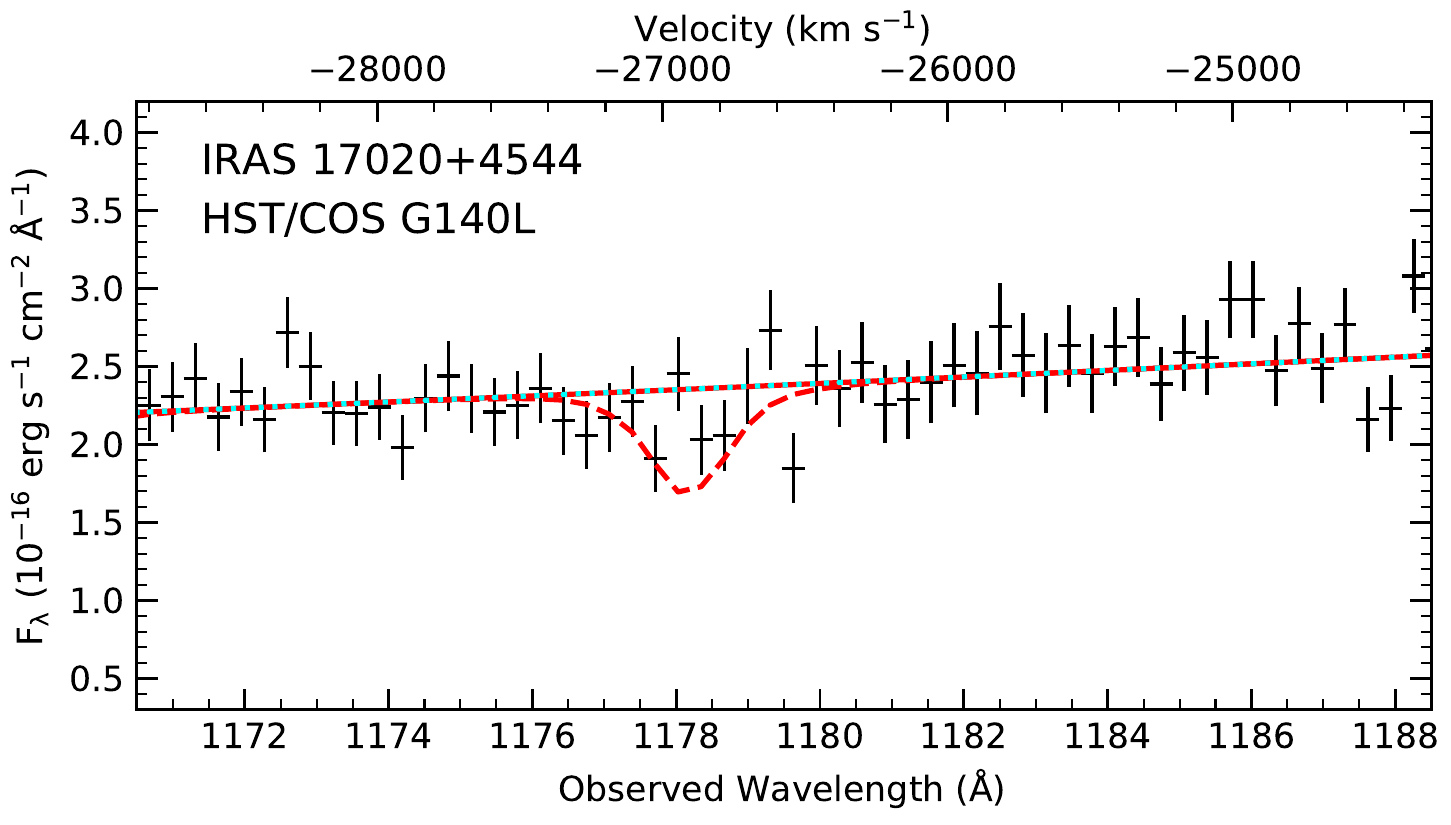}
}
\resizebox{\hsize}{!}{
\includegraphics[angle=0]{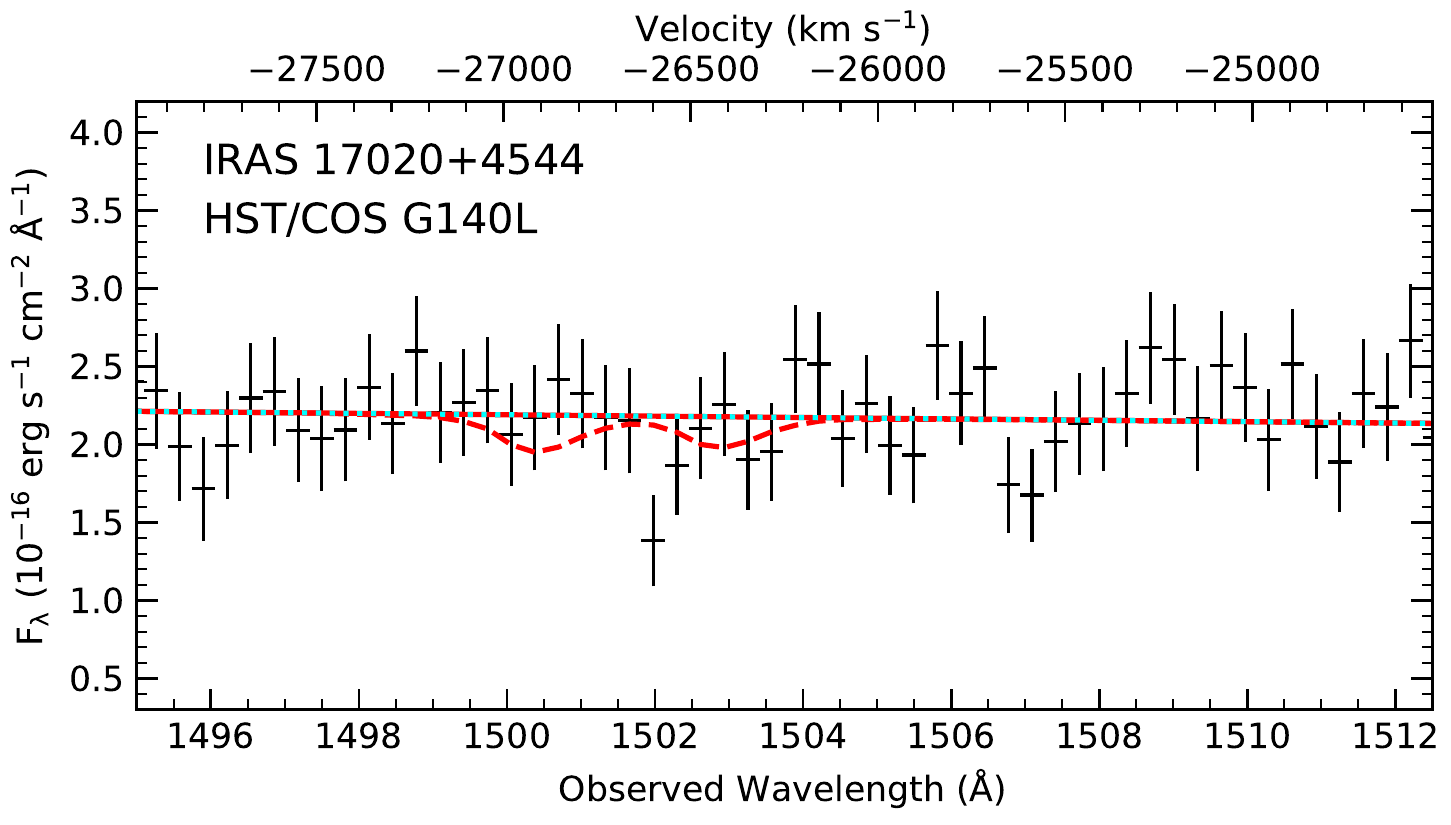}
}
\caption{Predicted \lya ({\it top panel}) and \ion{C}{4} ({\it bottom panel}) absorption lines for the high-velocity components of the X-ray UFO in \iras (UFO 1 and UFO 2 in Table \ref{table_xray}), plotted over the observed HST/COS spectrum. The low-ionization UFO 1 is shown in dashed red line. The highly-ionized UFO 2 (dotted red line) overlaps with the continuum (dotted cyan line) as it produces no UV absorption. The models are calculated using the \cloudy photoionization code, assuming full covering fraction, and are convolved with the LSF of COS/G140L. The COS data show no significant UV counterpart of the high-velocity X-ray UFO components (${v_{\rm out} = -26{,}900}$~\kms) is present in the COS spectrum, unlike the low-velocity UFO component (${v_{\rm out} = -23{,}430}$~\kms), which its UV counterpart through \lya absorption is found (Figure \ref{fig_ufo}). 
\label{fig_no_UFO}}
\end{figure}

Apart from the low-velocity components of the X-ray UFO, which we have modeled above, there are also two high-velocity components (UFO 1 and UFO 2 in Table \ref{table_xray}). These two components have an outflow velocity of about $-26{,}900$~\kms \citep{Sanf18}, and thus would fall in a different region of the COS spectrum. To check for the UV presence of these components, we calculated their model spectra with \cloudy using the parameters of \citet{Sanf18} given in Table \ref{table_xray}. We assumed a case of full-covering fraction. The predicted \cloudy models, which are blueshifted with ${v_{\rm out} = -26{,}900}$~\kms and convolved with the COS LSF, are plotted over the COS spectrum in Figure \ref{fig_no_UFO}. The COS data show no significant UV counterparts of these high-velocity X-ray UFO components are present in either the \lya region (top panel) or the \ion{C}{4} region (bottom panel). The model for UFO 2 shows that it is too highly ionized to produce any UV absorption line (Figure \ref{fig_no_UFO}), which is consistent with the COS data. The UFO 1 component produces some \lya absorption when assuming full covering fraction (Figure \ref{fig_no_UFO}). However, considering that this component may partially cover the UV source (such as in our Model B scenario), and hence the \lya absorption model could be even weaker than that displayed in Figure \ref{fig_no_UFO} (top panel) with ${\cf = 1}$, the predicted \cloudy model is consistent with the lack of significant detection in the COS spectrum.

If we adopt a scenario of a full-covering UV absorber (our Model A) the \NH that fits the \lya line matches that of the X-ray UFO 3 component. However, if both UFO 3 and UFO 4 components contribute to the \lya absorption (hence a higher \NH), then the partially-covering scenario (our Model B) becomes applicable. Nonetheless, whichever way the \lya feature is fitted, the other associated lines, such as the \ion{C}{4} and \ion{N}{5} doublets, are still too weak to be significantly detected in the COS spectrum according to our \cloudy calculations (see Figure \ref{fig_ref}, top right panel).

%

\subsection{Reasoning for the \lya explanation of the 1192 \AA\ feature}
\label{sect_ref}
\begin{figure*}
\centering
\resizebox{0.93\hsize}{!}{
\includegraphics[angle=0]{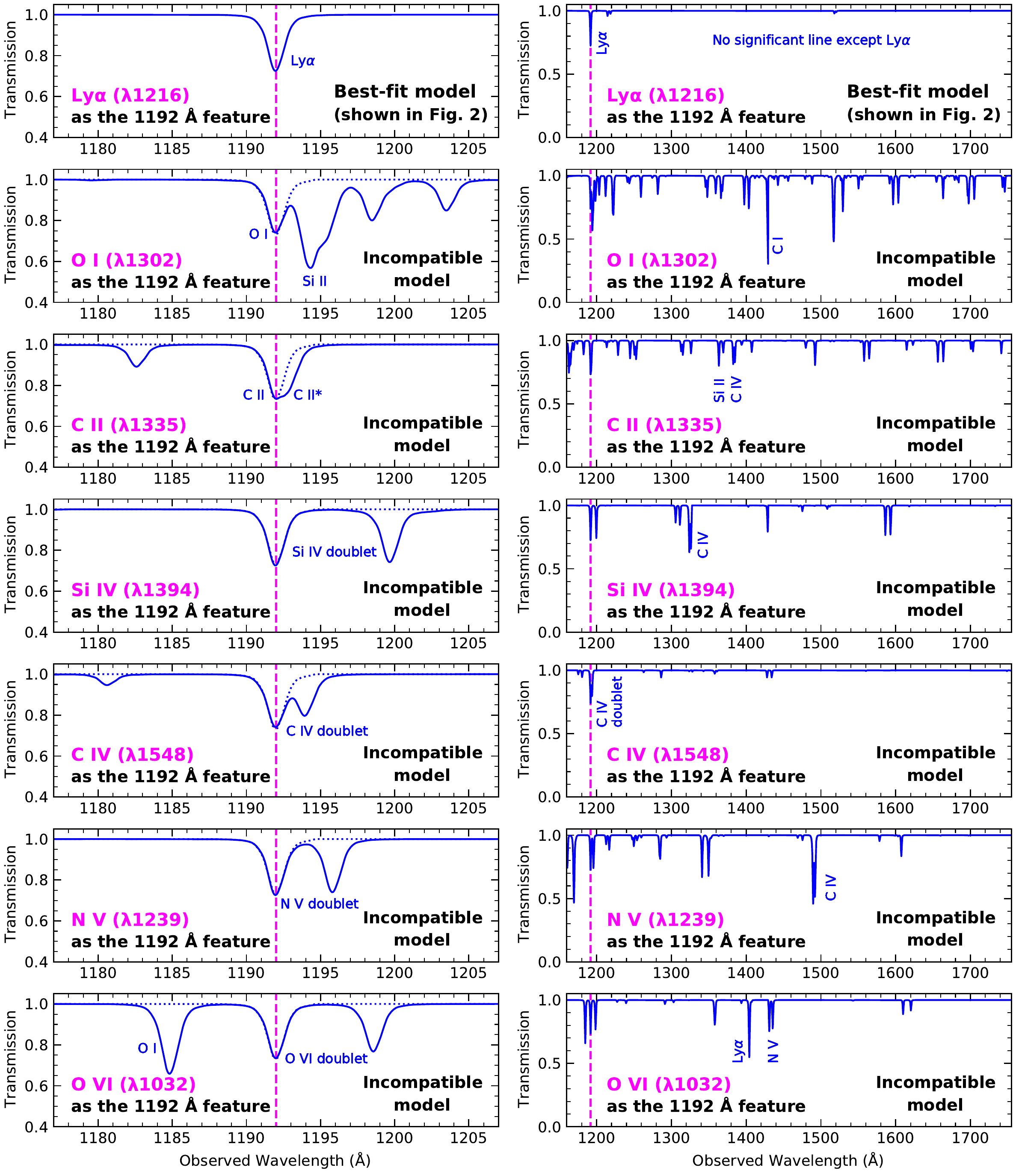}
}
\caption{Comparison of spectral models where the 1192 \AA\ feature is fitted with different ions. The photoionization and spectral modeling are carried out using the \cloudy code. The best-fit model, where the 1192 \AA\ feature is modeled as \lya, is taken from Figure \ref{fig_ufo} and displayed in the top panels. The left panels show close-ups of the region around 1192 \AA\ (like in Figure \ref{fig_ufo}) and the right panels show the models across the COS band (like in Figure \ref{fig_overview}). For reference the vertical dashed line in magenta marks the position of the 1192 \AA\ feature in all panels. The best-fit \lya model is superimposed as a blue dotted line on other models (left panels) to highlight differences between the models. All displayed model spectra are convolved with the LSF of COS/G140L, showing how the lines would be resolved with COS/G140L. Modeling the 1192 \AA\ feature with ions other than \lya would produce significant associated lines (either doublet of the same ion or lines from other ions), which are not seen in the COS spectrum (Figures \ref{fig_overview} and \ref{fig_ufo}). Some of the strongest of these associated lines have been labeled. In the case of the \lya model (top panels), the associated \ion{N}{5} and \ion{C}{4} doublets predicted by the model are too weak for detection in the COS spectrum, and also the predicted \ion{N}{5} doublet falls on the geocoronal \lya emission line at 1216 \AA\ and thus is unobservable.
\label{fig_ref}}
\end{figure*}

As mentioned earlier in Section \ref{sect_model}, the observed 1192 \AA\ feature (Figure \ref{fig_ufo}) can only be modeled as a \lya line. Here, we present spectral calculations using \cloudy that demonstrate this point. We modeled the 1192 \AA\ feature with different ions, from a wide range of ionization levels, that can produce absorption lines in the UV band: \ion{O}{1}, \ion{C}{2}, \ion{Si}{4}, \ion{C}{4}, \ion{N}{5}, and \ion{O}{6}. In each case, the modeling was carried out at the ionization parameter where each ion becomes most abundant, thus maximizing the potential of modeling the 1192 \AA\ feature with that ion. The results of our modeling are shown in Figure \ref{fig_ref}, where in the left panels close-ups of the 1192 \AA\ region are displayed, and on the right panels the models across the COS energy band are plotted. The top panels of Figure \ref{fig_ref} provide our best-fit model (from Figure \ref{fig_ufo}), where the 1192 \AA\ UFO feature is modeled as a \lya line. In this best-fit model the only significant UV line that is produced in the COS band is the \lya line (Figure \ref{fig_ref}, top panels). However, in the other models, shown in the subsequent panels of Figure \ref{fig_ref}, other significant absorption lines are also produced (either doublet of the same ion or lines from other ions), which are not seen in the COS spectrum (Figures \ref{fig_overview} and \ref{fig_ufo}).

The spectral models shown in Figure \ref{fig_ref} are convolved with the LSF of COS/G140L. In the case of ions that produce doublets (\ion{Si}{4}, \ion{C}{4}, \ion{N}{5}, and \ion{O}{6}), these lines are resolved and would be detected with COS/G140L. As shown in Figure \ref{fig_ref}, as a result of modeling the 1192 \AA\ feature with one of the doublet lines, the other line of the doublet is also produced, which is incompatible with the COS spectrum of \iras (Figure \ref{fig_ufo}, bottom left panel). Our spectral fitting presented in Figure \ref{fig_ufo} shows that the 1192 \AA\ feature is a single narrow line, thus modeling it with either \ion{Si}{4}, \ion{C}{4}, \ion{N}{5}, or \ion{O}{6} is not feasible. Furthermore, modeling the feature at 1192 \AA\ with other ions, such as \ion{O}{1} and \ion{C}{2}, results in the production of other associated lines from similarly ionized species, which are not present in the COS spectrum (Figure \ref{fig_overview}). For example, the \ion{O}{1} model also produces \ion{C}{1} and \ion{Si}{2} lines, and the \ion{C}{2} model makes \ion{Si}{2} and \ion{C}{4} lines. Similarly, the \ion{N}{5} and \ion{O}{6} models are accompanied by \ion{C}{4} and \lya/\ion{N}{5} absorption lines, respectively. We note that all the lines in the COS spectrum of \iras are identified (see Figure \ref{fig_overview}), and except the 1192 \AA\ feature, they are all either from the Galactic ISM or the warm absorber of the AGN.

In the case of ions that produce multiple absorption lines at different wavelengths in the UV band (\ion{O}{1} and \ion{C}{2}), we selected the ones that have the best chance of being compatible with the COS spectrum, so that their associated strong \lya line falls outside the detection range of COS. Still all these non-\lya models for the 1192 \AA\ feature are infeasible as described above. The ions that are selected for modeling the 1192 \AA\ feature in Figure \ref{fig_ref} are also representative of other similarly ionized species, and their absorption lines appear together in the UV band. We find that our results hold regardless of which ion/line is considered, and only the \lya model can fit the single narrow line at 1192 \AA\ without becoming incompatible with the COS spectrum.

%

Furthermore, in the case of some ions (\ion{O}{6}), not only other infeasible associated lines would be produced in the COS spectrum, the required spectral shift to model the 1192 \AA\ line would imply ultra-fast inflowing gas, which is likely not plausible. Also, in some cases such as the \ion{C}{4} model, the required blueshift would imply outflow velocities much greater than the outflow velocities of the X-ray UFOs in \iras. However, the primary reason the \ion{C}{4} model is rejected is that the only \ion{C}{4} line in the UV band is a doublet ($\lambda$1548 and $\lambda$1551), which is resolved by COS/G140L, whereas the 1192 \AA\ feature is a single narrow line (Figure \ref{fig_ufo}), making the \ion{C}{4} model incompatible with the COS data.

In summary, since (1) the non-\lya models for the 1192 \AA\ feature are incompatible with the COS spectrum, and (2) the \lya model is the one that is consistent with the low-velocity components of the X-ray UFO in terms of outflow velocity, ionization parameter, and column density, it is hence most reasonable to conclude that the UFO feature at 1192 \AA\ is a \lya absorption line. This is similar to the case of PG~1211+143 \citep{Kris18a}, where the UV spectral signature of the X-ray UFO is also only seen as a \lya line.
\section{Discussion and conclusions} 
\label{sect_discuss}
High-resolution UV spectroscopy complements X-ray observations of AGN, allowing a more detailed and complete picture of the ionization and kinematic structure of the outflows to be established. Our HST/COS spectroscopic study of \iras validates the X-ray UFO presence in this AGN. This UV UFO is seen in the far-UV band as a \lya absorption line at 1192.0 \AA, with a S/N of $6\, \sigma$, outflowing with a velocity of $-23{,}430$~\kms. The findings of our HST/COS investigation are consistent with those from previous X-ray studies of the UFO in \iras \citep{Long15,Sanf18}. Our modeling shows that the blueshifted \lya feature in the COS spectrum corresponds to the low-ionization, low-velocity component of the X-ray UFO. The UV spectral signature of the UFO is seen only through \lya absorption line because other associated absorption lines would be too weak to be significantly detected in the COS spectrum. This is according to the predictions from the photoionization modeling that we have carried out using the \cloudy code.

Our analysis of the COS spectrum of \iras shows that the other higher-velocity or higher-ionization components of the X-ray UFO (Table \ref{table_xray}) are not significantly detected in the COS spectrum. As the production of UV absorption lines is dependent on the ionization parameter of the gas, towards higher ionizations the column densities of the neutral and ionized UV species are greatly reduced, making their absorption lines too weak for detection. Also, the increased thermal line-broadening at higher ionizations makes the lines appear too shallow and blended with the continuum. In the case of the low-ionization, low-velocity component (${\log U = -0.4}$) that produces the \lya UFO feature, our \cloudy photoionization calculations show that this \lya line would have not been detectable in the COS spectrum if ${\log U > +0.3}$. Furthermore, components of the X-ray UFO may not fully cover the UV source (which is larger in size than the X-ray source), thus further weakening their UV absorption lines.

\iras stands out as one of only few AGN showing a truly multi-component UFO in terms of both ionization and velocity. Our finding of the \lya UFO feature makes \iras the first energy-conserving-outflow AGN with a UV counterpart. On the other hand, the UV counterpart of the UFO in PDS~456 \citep{Hama18} belongs to a momentum-conserving outflow \citep{Bisc19}. The multi-component nature of the UFO in \iras, and the relatively low \NH, $U$, and $v_{\rm out}$ of the \lya UFO component, points to a scenario where a `primary' powerful UFO has interacted with the ambient medium, entraining and shocking it, resulting in weaker `secondary' UFO components, such as the component seen in the UV. This may also explain why unlike broad UV absorption lines of powerful AGN winds, the \lya line in \iras is narrow. Such entrained-UFO explanations have been proposed in the past for multi-component and low-ionization UFOs \citep{Sanf18,Sera19,Long20}.

The profile of the UFO \lya line in the COS spectrum (Figure \ref{fig_ufo}) is consistent with instrumental broadening and any smaller thermal broadening which is produced by the \cloudy modeling. The narrowness of the UFO \lya line in \iras is in contrast to the \lya counterpart of the X-ray UFO in PG~1211+143 \citep{Kris18a}, which is a broad feature. This demonstrates that in searches for the UV counterpart of X-ray UFOs both narrow and broad features should be investigated. A recent study of a sample of quasars at high redshifts by \citet{Chart21} finds coexistence of intrinsic narrow UV absorption lines in quasars with X-ray UFOs. This is similar to the narrow UV absorption lines often associated with the X-ray warm absorbers \citep{Laha14}, which also have a considerable fraction of UFOs present. Multiphase outflows are thus seen in both low and high redshift AGN, which typically have different outflows in terms of their power and impact on their host galaxies. The multi-phase outflows, such as the one established in \iras, provide useful observational clues for linking the outflows at small scales near the black hole to the large scales of the host galaxy in a physically-consistent fashion.

Currently, X-ray spectroscopy of UFOs with CCD instruments leaves large uncertainties in the derived parameters of the photoionized gas. In particular, the ionization parameter and the outflow velocity cannot be constrained accurately as individual lines from different ionic species are blended together in the low-resolution CCD spectra. Also, high-resolution X-ray grating spectroscopy is often unfeasible due to the impractical amount of exposure time that is required. \iras is one of the few rare cases, where owing to its multi-component UFO and brightness, its components have been seen in the soft X-rays and the UV band. The upcoming high-resolution microcalorimeter X-ray missions, namely XRISM \citep{Tash18} and {\it Athena}/X-IFU \citep{Barr16}, will provide a major leap forward in X-ray studies of UFOs and overcome the shortcomings of the current missions. The tantalizing possibility of joint X-ray (XRISM) and UV (HST/COS) spectroscopy in the near future would open a new window in the UFO studies. They will facilitate all ionization and velocity components of the UFO to be mapped, from highly-ionized \ion{Fe}{26} to the neutral \ion{H}{1} \lya. Such X-ray/UV observations of UFOs, alongside theoretical predictions from hydrodynamical simulations of UFOs and their interaction with their environment, will advance our understanding of the UFO phenomenon in AGN.

\begin{acknowledgments}
This work was supported by NASA through a grant for HST program number 15239 from the Space Telescope Science Institute, which is operated by the Association of Universities for Research in Astronomy, Incorporated, under NASA contract NAS5-26555. A.L.L. acknowledges support from CONACyT grant CB-2016-286316. We thank the anonymous referee for providing constructive comments and suggestions that improved the paper.
\end{acknowledgments}
\facilities{HST (COS)}
\software{Cloudy \citep{Ferl17}}
\bibliography{references}{}
\bibliographystyle{aasjournal}
\end{document}